\documentclass[preprint,times]{elsarticle}

\usepackage{graphicx}
\usepackage{amssymb}
\usepackage{url}
\usepackage{textcomp}
\usepackage{booktabs}
\usepackage{array}
\usepackage{xcolor}
\usepackage{longtable}
\usepackage{listings}
\graphicspath{{./}}

\journal{SoftwareX}

\makeatletter
\def\ps@pprintTitle{%
  \let\@oddhead\@empty
  \let\@evenhead\@empty
  \def\@oddfoot{\footnotesize\itshape
    Published in SoftwareX 36 (2026) 103087 (CC BY 4.0)\hfill
    doi:10.1016/j.softx.2026.103087\upshape}%
  \let\@evenfoot\@oddfoot}
\makeatother

\newcolumntype{L}{>{\raggedright\arraybackslash}p{0.60\linewidth}}

\begin{document}

\begin{frontmatter}

\title{SecretSifter: A Runtime Secret-Detection Engine for Live Web Traffic\tnoteref{vor}}
\tnotetext[vor]{This is the version of record of an article published in
\emph{SoftwareX} \textbf{36} (2026) 103087, doi:
\url{https://doi.org/10.1016/j.softx.2026.103087}.
\textcopyright{} 2026 The Author. Published by Elsevier B.V. This is an open access
article under the CC BY 4.0 license
(\url{https://creativecommons.org/licenses/by/4.0/}). Please cite the published version.}

\author{Hemanth Gorijala}
\address{Independent Security Researcher \\ ORCID: 0009-0006-9810-4001}
\ead{gorijala2k16@gmail.com}

\begin{abstract}
Credentials embedded in client-delivered web artifacts, such as minified
JavaScript bundles, webpack chunks, server-side-rendered state objects, and JSON
or XML API responses, form a runtime attack surface that repository-time
(shift-left) secret scanners do not observe, because the credential enters the
served artifact at build time or at runtime, after every source-code gate has
passed. This paper presents SecretSifter, an open-source runtime secret-detection
engine that inspects live web traffic and reports secret-grade credentials at the
point they are delivered to the browser. The engine combines format-anchored and
context-gated detection rules, semantic key-name extraction, Shannon-entropy
scoring, and dedicated detection of hardcoded CryptoJS/AES material across the
full served surface. SecretSifter ships as three editions over one shared
detection core: a Burp Suite extension, a Chrome extension, and a standalone
Windows and macOS desktop application, with a Bearer-authenticated REST and MCP
interface for automation. On a 198-value benchmark of real production
client-exposed values (GT-198), the deterministic detection stage reached 79.8\%
recall at 86.8\% precision in default configuration, the highest F1 of any system
evaluated. The detection core is released under an open-source (MIT) license, and
a de-identified reproducibility bundle and a public synthetic demo target are
provided.
\end{abstract}


\end{frontmatter}

%% ======================= Required metadata =======================
\section*{Code metadata}
{\footnotesize
\begin{longtable}{@{}p{0.03\linewidth}L@{}}
\toprule
\multicolumn{2}{@{}l}{\textbf{Code metadata}}\\
\midrule
\endhead
C1 & Current released version: Burp extension v1.0.3; AI/MCP variant v1.0.2; Chrome extension v1.1.2; desktop app v1.7.7\\
C2 & Permanent link to code/repository (evaluated versions pinned by release tag and commit):\newline
     Source (Burp extension, full detection core): \url{https://github.com/secretsifter/secretsifter-burp}, tag v1.0.3 (commit \texttt{dc40c602bee55e9231245c7807dcbc12432ba552})\newline
     Source (AI/MCP variant, same core plus triage and MCP server): \url{https://github.com/secretsifter/secretsifter-burp-mcp}, tag v1.0.2 (commit \texttt{89639907697010a27cdb042ca84be1752c50dc30})\newline
     Distributions of the same core: Chrome extension v1.1.2 via Web Store (\url{https://chromewebstore.google.com/detail/secretsifter-live-credent/okjoofmgkkgopigaipccailblmpcchjo}); Windows/macOS desktop app tag v1.7.7 (commit \texttt{2cd792f0c77503e1390c321ebe689eb4aa6eab4e}) via signed installers (\url{https://github.com/secretsifter/secretsifter-desktop})\newline
     Archived, de-identified reproducibility bundle (per-value detection matrix and significance results): \url{https://doi.org/10.5281/zenodo.21695364}\newline
     Evaluation workflow in the source repository (de-identified scoring script, detection matrix, six-rule overlay, and per-tool commands): \url{https://github.com/secretsifter/secretsifter-burp/tree/main/evaluation}\\
C3 & Reproducible target: InsecureShield synthetic demo application and scan
     configuration, \url{https://github.com/secretsifter/insecureshield-demo}.
     Synthetic credentials only.\\
C4 & Legal code license: MIT for the open-source detection core and its Burp and
     AI/MCP extensions (the repositories in C2 on which the evaluation is run); the
     Chrome extension and desktop application are distributed as packaged builds of
     that same core under a proprietary end-user license.\\
C5 & Code versioning system: git\\
C6 & Languages, tools, and dependencies (per edition).
     \emph{Burp extension and AI/MCP variant (the open-source core):} Java~17;
     Montoya API~2026.2 (compile-time only, provided by Burp at runtime);
     Gson~2.10.1; JUnit~5.10.2 (test only). AI-assisted triage and the MCP server
     use the JDK built-in \texttt{java.net.http} client (no third-party HTTP
     dependency).
     \emph{Desktop application:} Java~17; Swing with FlatLaf~3.4; Microsoft
     Playwright for Java~1.45.0; Bouncy Castle \texttt{bcpkix-jdk18on}~1.78.1;
     \texttt{org.json}~20231013; an embedded HTTP MITM proxy.
     \emph{Chrome/Firefox extension:} JavaScript, Manifest~V3 (ES modules with a
     module service worker), using the Chrome DevTools Protocol via the
     \texttt{debugger} API and a MAIN-world content script.\\
C7 & Build and runtime requirements (per edition).
     \emph{Burp extension and AI/MCP variant:} JDK~17 and Gradle~8.9 with the
     Shadow plugin~8.3.5 (\texttt{./gradlew shadowJar}); the extension loads into
     Burp Suite~2026.2 or later, and automated passive scanning requires Burp
     Suite Professional (live proxy-based detection also runs in the Community
     edition).
     \emph{Desktop application:} JDK~17 and Gradle~8.7 (Shadow~8.3.5) to build,
     packaged with \texttt{jpackage} and WiX Toolset~3.11+ into a signed Windows
     installer that bundles an OpenJDK~21 runtime, so no separate JRE is required
     to run it. At runtime the Playwright crawler drives a Chromium-based browser
     (a bundled Chromium is used as a fallback, or a system Chrome/Edge install),
     and the embedded MITM proxy requires the operator to trust the tool's
     locally generated Bouncy Castle CA certificate for TLS interception.
     \emph{Chrome/Firefox extension:} no build step (raw ES modules); requires a
     Chromium-based browser that supports Manifest~V3 with MAIN-world content
     scripts, i.e.\ Chrome/Chromium~111 or later.
     \emph{Optional AI triage (all editions):} the Burp-native back end uses
     Burp's AI; the remote back ends additionally require an external
     LLM endpoint (an OpenAI-compatible or Anthropic URL, or a self-hosted model
     served through Ollama or LM~Studio) and, for hosted providers, an API key
     supplied by the operator. Triage is optional and off by default, so these are
     not required to run any edition's detection.\\
C8 & Developer documentation: repo \texttt{docs/} (Documentation and Engineering
     Reference, PDF/HTML) and README\\
C9 & Support email: gorijala2k16@gmail.com\\
\bottomrule
\end{longtable}
\addtocounter{table}{-1}%
}

%% ======================= 1 =======================
\section{Motivation and significance}
Modern software authenticates to cloud services, databases, and third-party APIs
using secrets such as API keys, access tokens, and passwords. When one of these
secrets is exposed where an untrusted party can read it, an attacker can reuse it
directly, so finding exposed secrets automatically, a task known as automated
secret detection, is a core software-security activity. Automated secret
detection is today dominated by shift-left tooling: scanners that run
over source repositories and CI pipelines, such as TruffleHog~\cite{trufflehog},
git-secrets~\cite{gitsecrets}, detect-secrets~\cite{detectsecrets}, and SAST
rules. These tools are effective on the surface they observe, committed source,
where secret leakage is a well-documented problem~\cite{meli2019}. They are
structurally blind, however, to credentials that reach production without ever
appearing in a repository. Two mechanisms produce this blind spot. The first is
build-time static substitution. Bundlers such as webpack~\cite{webpack},
Vite~\cite{vite}, esbuild~\cite{esbuild}, and the Angular CLI compile environment
values directly into the output bundle, so the credential exists only in the
deployed artifact. The second
is runtime delivery. Configuration APIs and server-side-rendered state objects
serve credentials to the browser on demand, again bypassing any repository scan.

The companion measurement study~\cite{gorijala2026} quantifies this gap on an
enterprise corpus of 113 applications and shows that a substantial fraction of
production credentials are recoverable only from served content. This paper
evaluates SecretSifter on its own 198-value benchmark of client-exposed values
(GT-198), introduced in Section~\ref{sec:impact}. Closing the gap calls for a tool
that intercepts live traffic, normalizes heterogeneous served artifacts, and
applies detection tuned for the noise profile of minified production JavaScript.
Existing runtime-adjacent tools are narrow. File-oriented JavaScript scanners
miss HTTP responses and headers; HTTP-template scanners miss file content; none
cover split chunks, SSR blobs, JSON and XML bodies, and outbound headers in a
single pass. SecretSifter targets this gap. It combines runtime interception,
full-surface coverage, context-aware detection, and optional LLM-assisted triage
in one self-hosted tool released under an open-source license.

The contribution of this paper is the software: the shared detection core, the
three editions, and the automation interface. It is complementary to a companion
study published in IEEE Access~\cite{gorijala2026}: that paper investigates
runtime credential exposure in deployed web applications, whereas the present
paper focuses on the reusable SecretSifter software, namely its architecture, its
three released editions, its code metadata, and its reproducibility package. The
GT-198 evaluation provides engineering evidence for the software.

%% ======================= 2 =======================
\section{Software description}

\subsection{Software architecture}
\begin{figure*}[t]
  \centering
  \includegraphics[width=0.92\textwidth]{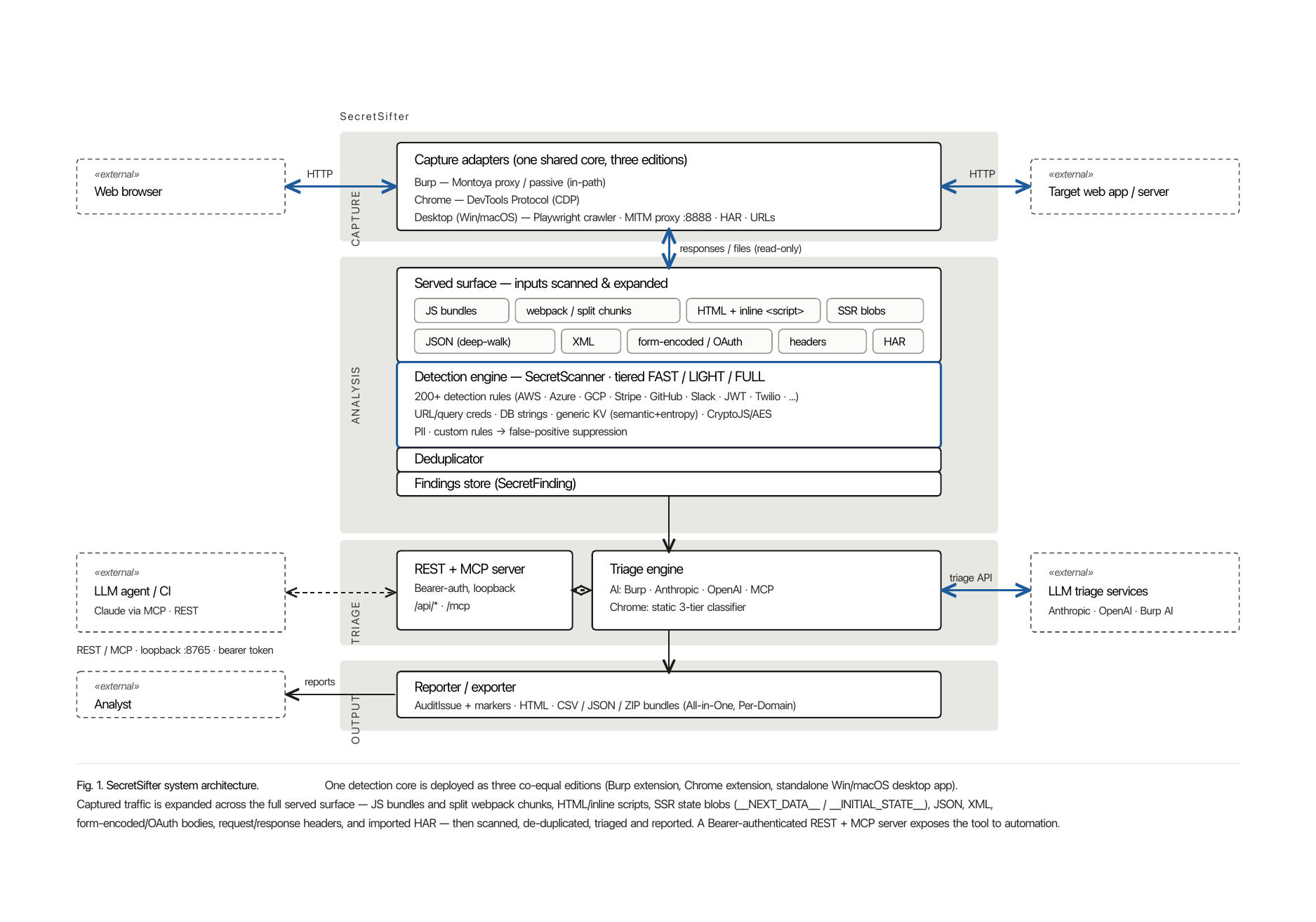}
  \caption{SecretSifter system architecture. One detection core is deployed as
  three editions (Burp extension, Chrome extension, and a standalone Windows and
  macOS desktop application). Captured traffic is expanded across the full served
  surface, including JavaScript bundles and split webpack chunks, HTML and inline
  scripts, SSR state blobs (\texttt{\_\_NEXT\_DATA\_\_},
  \texttt{\_\_INITIAL\_STATE\_\_}), JSON, XML, form-encoded and OAuth bodies,
  request and response headers, and imported HAR. Content is then scanned,
  de-duplicated, triaged, and reported. A Bearer-authenticated REST and MCP
  server exposes the tool to automation.}
  \label{fig:arch}
\end{figure*}

The architecture is shown in Figure~\ref{fig:arch}. SecretSifter is one detection
core deployed as three editions that differ only in how they capture traffic and
surface findings. A single detection model that covers proxy, in-browser, and
standalone deployment is the central design decision of the tool.

\paragraph{Burp Suite edition} The extension registers through the Montoya
API~\cite{montoya}. \texttt{SecretScannerExtension} implements
\texttt{BurpExtension} and, in \texttt{initialize(MontoyaApi)}, wires a
thread-safe configuration model (\texttt{ScanSettings}), a stateless detection
engine (\texttt{SecretScanner}), and three interception paths: a proxy response
handler (\texttt{SecretProxyHandler}) that fires on all traffic, an optional
passive scan check (\texttt{SecretScanCheck}, Burp Pro), and a background
sitemap-sweep thread that retroactively scans existing site-map entries on load.
Findings are grouped and de-duplicated (\texttt{SitemapDeduplicator}), reported
as Montoya \texttt{AuditIssue} objects, optionally triaged
(\texttt{AiTriageProvider}), and exportable as a self-contained HTML report
(\texttt{HtmlReportGenerator}). A combined settings and bulk-scan UI is
registered as a suite tab, and an MCP bridge is included in the full build
variant. Because the BApp Store prohibits extensions that open listening sockets,
the Burp edition is built from one codebase as two artifacts: a store build with
the MCP bridge excluded at compile time, and a full build (repository
\texttt{secretsifter-burp-mcp}) that adds a loopback JSON-RPC MCP server for
cooperative triage. These correspond to the two source repositories listed in C2.

\paragraph{Chrome edition} A Manifest V3 extension attaches via the Chrome
DevTools Protocol~\cite{cdp} (\texttt{chrome.debugger}) and enables the
\texttt{Network} domain to intercept responses and fetch bodies for JavaScript,
JSON, XML, and HTML. The detection library and entropy math are ported to
JavaScript. Classification uses a static, no-network three-tier classifier
(\texttt{classifier.js}) rather than an LLM. UI surfaces include a popup, an
options page, a DevTools panel, and an in-page report.

\paragraph{Desktop edition (Windows and macOS)} A standalone application (Java and
Swing, packaged with jpackage, no Burp licence required) acquires traffic through
four modes: a Playwright-driven real-Chrome crawler that masks
automation flags so the browser stays compatible with bot-management systems (for
example, Cloudflare and Akamai) during authorized testing; an embedded HTTP MITM proxy
(\texttt{127.0.0.1:8888}, per-host TLS via a Bouncy Castle CA); DevTools HAR
import; and bulk-URL scanning. It runs the same \texttt{SecretScanner} core and
rule library, presents a findings dashboard with a request and response details
panel, produces All-in-One and Per-Domain report bundles (HTML, CSV, JSON), and
exposes a Bearer-authenticated REST and MCP server (\texttt{McpServer}).

All three editions share one detection engine (\texttt{SecretScanner}) and rule
library. They diverge only at capture (proxy, CDP, or crawler with MITM and HAR),
at triage (LLM-assisted in Burp and Desktop, static classifier in Chrome), and at
output. The open-source release (C2) is the Java detection core and its Burp
capture layer, in the two repositories \texttt{secretsifter-burp} and
\texttt{secretsifter-burp-mcp}; this is the code on which the evaluation in
Section~\ref{sec:impact} is run. The Chrome and desktop editions are distributed
as packaged builds of the same core (Web Store listing and signed installers) and
are the JavaScript and Swing ports of that engine; their edition-specific capture
and UI code is not part of the open repositories.

\subsection{Software functionalities}
\label{sec:func}
\paragraph{Detection pipeline} Content is intercepted without modification,
deep-copied off the proxy thread to avoid buffer recycling, and dispatched by
content type through three configurable scan tiers (FAST, LIGHT, FULL). Detection
methods are: (i) a pattern library of 134 format-anchored vendor rules (GitHub, GitLab, AWS,
Stripe, OpenAI, Anthropic, Slack, PEM keys, and others) and 52 context-gated
rules that require a vendor keyword near the value (Azure APIM, Twilio, Snowflake,
and others); (ii) URL
and query-embedded credential detection; (iii) database connection-string
detection; (iv) context-gated generic key-value extraction with a false-positive
suppression cascade covering framework directives, CSS-module hashes, OIDs,
reCAPTCHA keys, and hex strings without crypto context; (v) Shannon-entropy
scoring~\cite{shannon1948} (\texttt{shannonEntropy}, default threshold 3.5);
(vi) dedicated detection of hardcoded CryptoJS/AES~\cite{cryptojs} passphrases,
keys, and ciphertext, the credential class that pattern-only tools do not flag;
and (vii) user-defined custom rules, including a raw custom-rules-only mode.
Cross-cutting suppression handles JWT recognition, opaque-bearer filtering, CDN
blocklisting, and cross-request de-duplication.

\paragraph{Runtime coverage} Beyond initial JavaScript bundles, SecretSifter
follows split webpack and Next.js chunks, extracts SSR state blobs
(\texttt{\_\_NEXT\_DATA\_\_}, \texttt{window.\_\_CONFIG\_\_}-style objects, and
\texttt{<script type="application/json">} tags), recurses into inline HTML
scripts, deep-walks JSON bodies with a bounded-depth Gson traversal, inspects XML
leaves and form-encoded OAuth token responses, and, when enabled, scans outbound
request headers and bodies, raising the severity of detected token-issuance
flows.

\paragraph{AI triage} \texttt{AiTriageProvider} supports four back ends: Burp
native AI, Anthropic (\texttt{claude-opus-4-7}), OpenAI-compatible endpoints, and
a cooperative MCP mode~\cite{mcp} in which an external LLM agent performs triage.
Each finding is classified as real or false-positive, after which deterministic
vendor rules, not the model, set the final severity, so the risk rating is
reproducible and does not vary with model output. Because the OpenAI-compatible
back end accepts any endpoint implementing that API, triage can also be driven by a
self-hosted or local model (for example, a Qwen model served through Ollama or LM
Studio), letting an operator run triage entirely on the local machine with no data
leaving it. Triage is optional and disabled by default, and no data leaves the machine
unless the operator configures a remote back end and invokes it. When a remote
back end is selected, SecretSifter sends only a compact record of each finding to
the configured endpoint --- the rule and key name, the matched value (truncated to
a short prefix once it exceeds forty characters), a bounded window of surrounding
context, and the source URL with its query string removed --- never the full
response body or page, and it retains no findings beyond the session. For
engagements with data-handling constraints the self-hosted (local model) or
Burp-native back end is recommended, keeping all finding data on the operator's
machine.

\paragraph{Automation interface (REST and MCP)} The Burp edition's
\texttt{McpBridge} exposes a JSON-RPC 2.0 server on loopback
(\texttt{127.0.0.1:8765}, path \texttt{/mcp}). The Desktop edition's
\texttt{McpServer} additionally serves a Bearer-authenticated REST API
(\texttt{/api/scan}, \texttt{/api/findings}, \texttt{/api/report},
\texttt{/api/settings}, \texttt{/api/rules}, \texttt{/api/logs}) alongside
\texttt{/mcp}. Both are disabled by default and protected by a regenerable
per-instance bearer token. They publish operations to start and cancel scans,
query status and findings, edit findings and settings, manage custom rules,
export reports, and drive cooperative triage, which lets an external LLM agent or
CI pipeline operate the tool programmatically.

\paragraph{Configuration and output} \texttt{ScanSettings} exposes the scan tier,
entropy threshold, PII scanning, request scanning, custom rules, AI provider and
keys, and CDN and key allow/block lists. Settings persist per edition (Burp
preferences, Chrome storage, or a Desktop config file). Findings
(\texttt{SecretFinding} records) carry the rule, key name, matched value,
severity, confidence, line number, context, and source and target URLs. Output
is edition-appropriate: Burp \texttt{AuditIssue} entries with in-response byte
markers, an in-page report in Chrome, and a findings dashboard with report
bundles in Desktop. All editions share a self-contained HTML report with severity
summaries, filtering, and embedded CSV and JSON export.

%% ======================= 3 =======================
\section{Illustrative examples}
\label{sec:examples}
This section demonstrates the tool on a public, synthetic target so the
walk-through is reproducible and exposes no real secrets. It is distinct from the
quantitative evaluation in Section~\ref{sec:impact}, which measures accuracy on
real production credentials that cannot be redistributed. The examples run
against InsecureShield, a purpose-built vulnerable demo application (a mock
insurance client portal) that hosts only synthetic credentials. The demo target
and scan configuration are public (see the Code metadata table, entry~C3).

\paragraph{Reproduction} The walk-through runs end to end from the public
repository: \texttt{git clone} the InsecureShield demo, then \texttt{npm install}
and \texttt{npm start}; the portal serves at \url{http://localhost:3000}. Point a
SecretSifter edition at it: in the Burp edition, place the local portal in scope
and run a bulk scan from the SecretSifter tab; in the Chrome edition, load the
MV3 build and browse the portal with capture enabled. The bulk scan reports 29
HIGH findings, including the APIM subscription key and the
\texttt{U2FsdGVkX1\dots} CryptoJS blob. The repository README lists the exact
endpoints and the per-file credential inventory.

\begin{figure}[t]
  \centering
  \includegraphics[width=\linewidth]{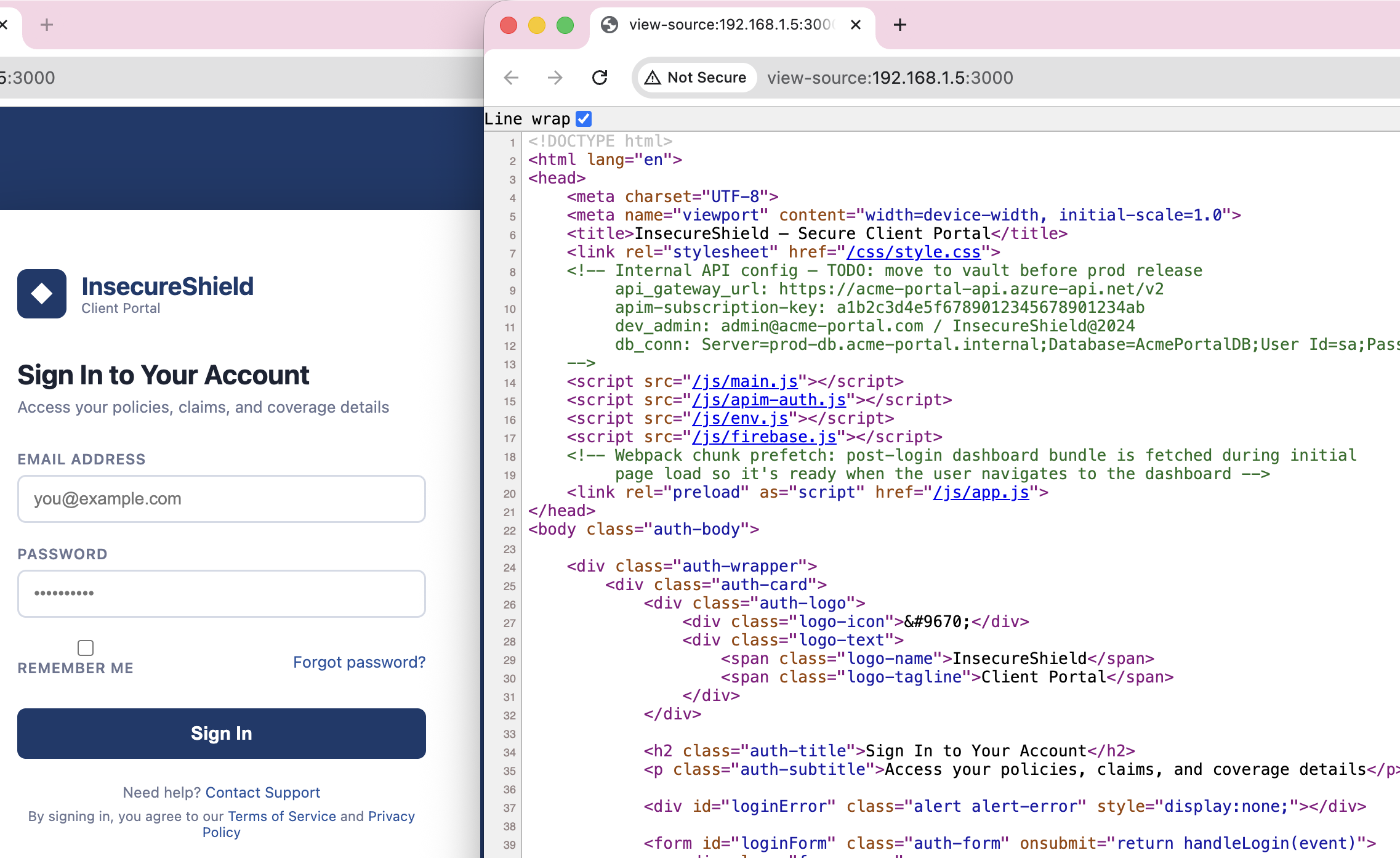}
  \caption{InsecureShield demo target. Secret-grade credentials are served in an
  HTML comment and in linked scripts, the runtime surface that repository
  scanners never observe.}
  \label{fig:target}
\end{figure}

\paragraph{Setup (Figure~\ref{fig:target})} The InsecureShield portal serves an
APIM subscription key, an Azure AD-style credential, a database connection
string, and admin credentials directly in an HTML comment and across several
linked scripts (\texttt{main.js}, \texttt{apim-auth.js}, \texttt{env.js},
\texttt{firebase.js}) plus a prefetched webpack chunk (\texttt{app.js}). This is
the runtime surface a repository scan never sees.

\begin{figure}[t]
  \centering
  \includegraphics[width=\linewidth]{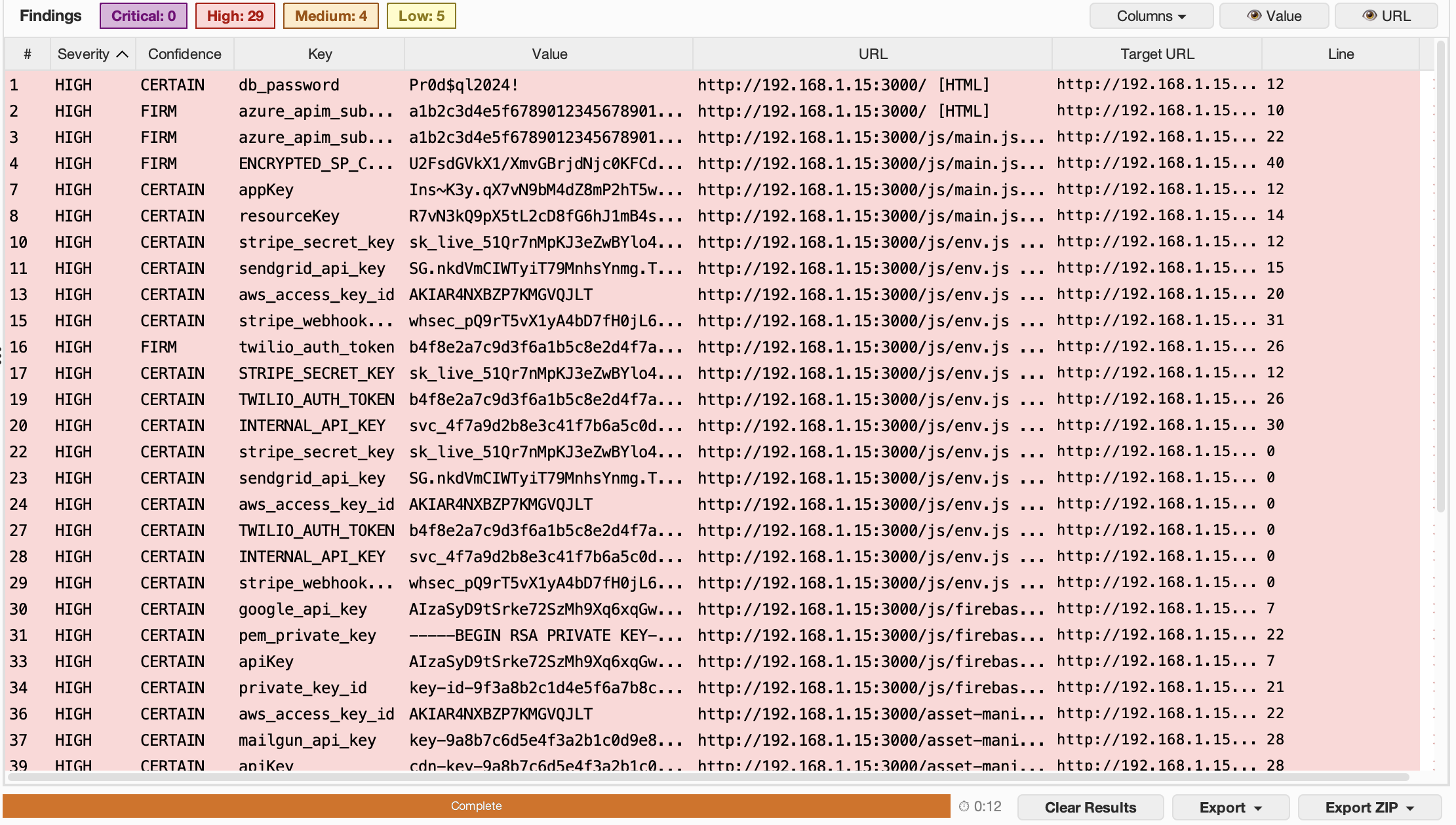}
  \caption{Bulk-scan findings. An Azure APIM subscription key (Example A) and a
  CryptoJS/AES blob (Example B) recovered from the served surface, among 29 HIGH
  findings.}
  \label{fig:bulk}
\end{figure}

\paragraph{Examples A and B (Figure~\ref{fig:bulk})} A single SecretSifter bulk
scan of the target surfaces 29 HIGH findings across the served surface. Two are
representative. Example A is an Azure APIM subscription key recovered from both
the HTML and the served \texttt{main.js} bundle. Example B is an
\texttt{ENCRYPTED\_SP\_C\dots} CryptoJS/AES blob (\texttt{U2FsdGVkX1\dots}, the
base64 \texttt{Salted\_\_} marker) that pattern-only scanners do not classify.
The scan also recovers Stripe, SendGrid, AWS, Twilio, Google, and PEM-key
material across \texttt{env.js}, \texttt{firebase.js}, and the asset manifest.

\paragraph{Example C (Listing~\ref{lst:mcp})} With the MCP bridge enabled, an
external LLM agent performs cooperative triage. It calls
\texttt{secretsifter\_get\_triage\_request}, classifies the finding batch as real
or false-positive, and returns verdicts via
\texttt{secretsifter\_submit\_triage\_results}, which update the severities and
the exported report.

\begin{figure}[t]
  \centering
  \includegraphics[width=\linewidth]{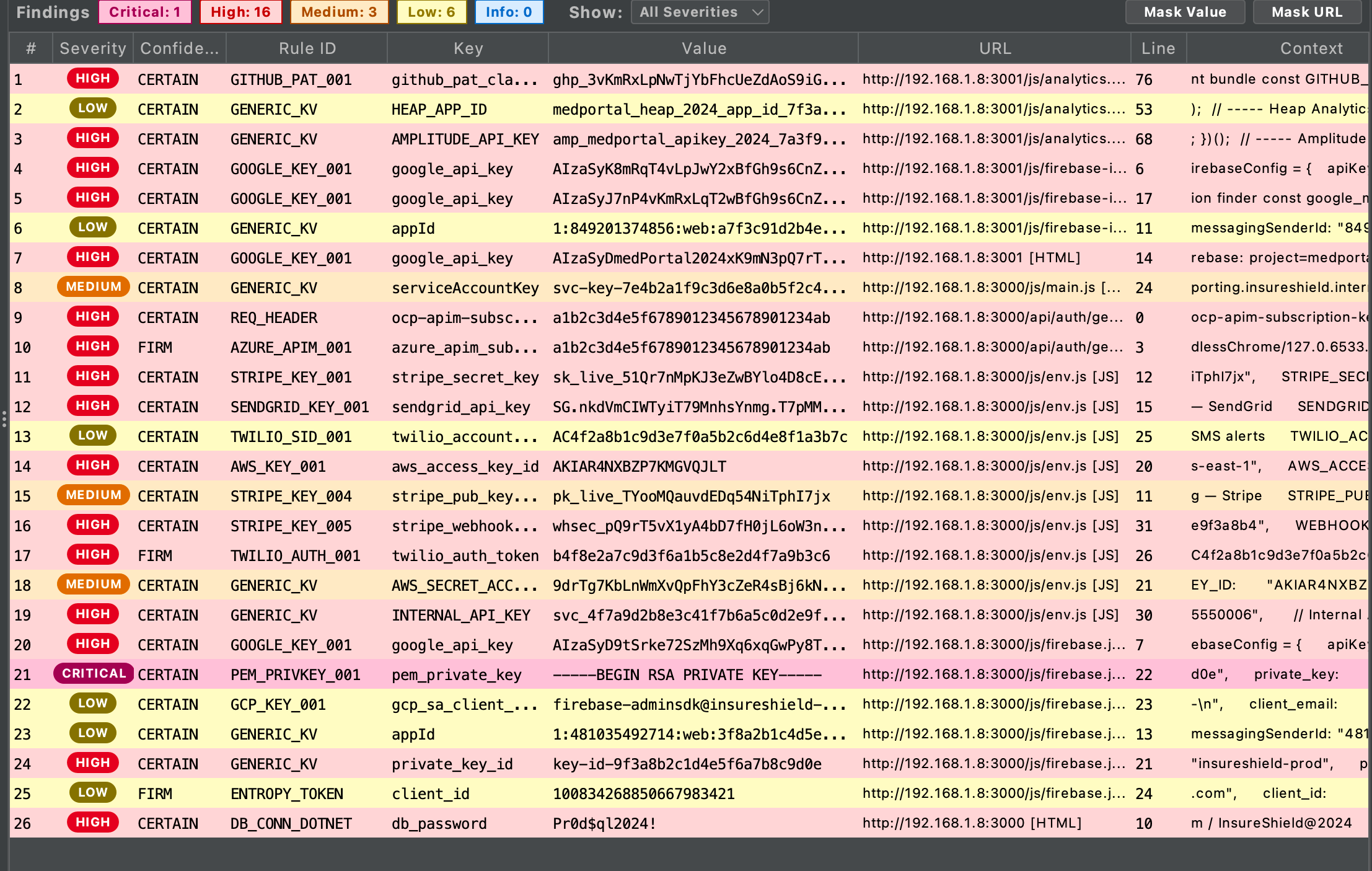}
  \caption{Desktop edition (Windows and macOS). Findings dashboard from a
  Browser-Mode crawl, with HAR import and report bundles, no Burp licence
  required.}
  \label{fig:desktop}
\end{figure}

\paragraph{Desktop edition (Figure~\ref{fig:desktop})} The standalone Windows and
macOS application scans the same class of target with Browser Mode (the
Playwright crawler) enabled and no Burp licence required. It surfaces the full
findings set (a GitHub PAT, Google and Firebase keys, an Azure APIM subscription
key, Stripe, SendGrid, Twilio, and AWS credentials, a PEM private key, and a
.NET database password) in its findings dashboard, alongside HAR import and
All-in-One and Per-Domain report bundles.

\begin{lstlisting}[float=t,caption={SecretSifter MCP server configuration. This
configuration enables a loopback, Bearer-authenticated MCP endpoint through which
an external LLM agent can perform cooperative triage using the SecretSifter MCP
tools.},label={lst:mcp}]
{
  "mcpServers": {
    "secretsifter": {
      "type": "http",
      "url": "http://127.0.0.1:8765/mcp",
      "headers": {
        "Authorization": "Bearer <token>"
      }
    }
  }
}
\end{lstlisting}

%% ======================= 4 =======================
\section{Impact}
\label{sec:impact}

\subsection{Benchmark evaluation}
SecretSifter is evaluated on GT-198, this paper's benchmark of client-exposed
values, used here for a software-engineering comparison. GT-198 holds 198 scored values recovered from
production JavaScript across 56 benchmark applications drawn from a single
authorized engagement; of 206 candidates originally extracted, adjudication
reclassified eight as non-credentials. It partitions into 161 static values
present as bundle text and 37 CryptoJS-AES encrypted-configuration blobs whose
plaintext requires decryption with a co-located key (Table~\ref{tab:types}). The
static base was anchored by a Claude Opus 4.7 oracle with no access to any
evaluated scanner, each candidate independently cross-validated by a second-vendor
model (GPT-5.5, 89.9\% high-confidence confirmation) and then extended by a manual
analyst pass that added values the models did not surface; because the ground
truth is single-class, the cross-vendor figure is reported as a confirmation rate
rather than a chance-corrected $\kappa$. Recall is measured on SecretSifter's
deterministic detection stage (anchored patterns, entropy scoring, and
decryptors), independent of the optional LLM triage layer, so the figures reflect
the detection engine rather than any model shared with the labelling process. All
recall figures below are measured on the real GT-198 values, not on the synthetic
demo of Section~\ref{sec:examples}.

The positive class includes 40 OAuth App\_ID, client\_id, and resource-URI values
that are not secrets in isolation (Table~\ref{tab:types}). These are retained
because they are chain-completing identifiers: when co-located with an exposed
secret they are required to determine whether that secret can be exercised in
practice, so they are part of the client-side exposure a runtime detector must
surface. Excluding these 40 identifiers would leave 158 values: 121 static values
and 37 encrypted-configuration values. We have not recomputed the comparative
metrics for this reduced set. The CryptoJS-AES encrypted-configuration class (37
values recovered by no other out-of-the-box scanner), which is the chief source of
SecretSifter's advantage, is disjoint from these 40 identifiers and is therefore
unaffected by their exclusion.

\begin{table}[t]
\footnotesize
\centering
\caption{GT-198 composition: 198 client-exposed values (161 static plus 37
CryptoJS encrypted-configuration), across 56 benchmark applications. OAuth App\_ID
and resource URI are chain-completing identifiers, not secrets in isolation;
excluding those 40 gives a secrets-only static base of 121.}
\label{tab:types}
\begin{tabular}{@{}lr@{}}
\toprule
Value type & Count \\
\midrule
\multicolumn{2}{@{}l}{\emph{Static (161)}}\\
APIM subscription key & 51 \\
Azure AD client\_secret & 33 \\
OAuth App\_ID / client\_id & 26 \\
OAuth resource URI & 14 \\
Other API key/token (incl.\ App Insights iKey) & 12 \\
JWT & 11 \\
Plaintext user credential / client\_secret pair & 7 \\
CyberArk AIM & 4 \\
APIM key by shape & 3 \\
\midrule
\multicolumn{2}{@{}l}{\emph{Encrypted configuration (37)}}\\
CryptoJS-AES \texttt{Salted\_\_} blob & 37 \\
\midrule
\textbf{Total} & \textbf{198} \\
\bottomrule
\end{tabular}
\end{table}

Ten systems (nine production scanners and one LLM oracle) were scored against
GT-198 in their default, out-of-the-box configuration. Alongside SecretSifter, the production
scanners are TruffleHog~\cite{trufflehog}, SecretFinder~\cite{secretfinder},
JSluice~\cite{jsluice}, Titus~\cite{titus}, JSMiner~\cite{jsminer},
Nuclei~\cite{nuclei}, Sensitive Discoverer~\cite{sensitivediscoverer}, and
Cariddi~\cite{cariddi}; per-tool versions are recorded in the extended
evaluation report~\cite{gorijala2026zenodo}.

SecretSifter runs its complete detection engine
unchanged and is configuration-invariant because it ships its full rule set by
default. That report also presents a tightened static-subset
analysis in which the four rule-extensible baselines receive a common six-rule
overlay (Azure client ID, tilde-format client secret, 32-character APIM key, JWT,
App Insights instrumentation key, and CryptoJS blob); on static text under that
overlay the strongest static scanners become statistically indistinguishable from
SecretSifter (Holm-corrected McNemar), so the software's advantage on this corpus
rests on its out-of-the-box coverage, its decryption-aware handling of the
CryptoJS class, and its noise suppression rather than on a claimed static-text
superiority.

Table~\ref{tab:results} reports, for each system scored over GT-198
in default configuration, the values detected (true positives), the
non-credential strings emitted (false positives), and the resulting precision,
recall, and F1. A false
positive here is a non-credential string that survives the tool's own filtering
and reaches its final findings list, that is, noise the user actually sees, not a
raw pattern match. Precision, recall, and F1 are computed on GT-198 in the standard way. The nine
production scanners are the field under test; the tenth entry, the Claude Opus
oracle, anchored the ground-truth extraction and is listed only as a labelling
reference. The oracle recovers the 161 static values by construction but none of
the 37 CryptoJS plaintext values from direct bundle reads, so its recall is
81.3\% rather than 100\%.

\begin{table}[t]
\footnotesize
\centering
\caption{Detection performance on the GT-198 corpus of real production values,
default (out-of-the-box) configuration. Det.\ = values detected (true positives);
FP = non-credential strings that survive each tool's own filtering and reach its
final findings list; precision $=$ Det/(Det+FP), recall $=$ Det/198, F1 $=$
$2PR/(P+R)$. The Claude Opus oracle anchored the ground-truth extraction and is
shown for reference, not as a competing scanner.}
\label{tab:results}
\begin{tabular}{@{}lrrrrr@{}}
\toprule
Scanner & Det. & FP & Prec. & Rec. & F1 \\
\midrule
\multicolumn{6}{@{}l}{\emph{Production scanners (field under test)}}\\
\textbf{SecretSifter} & \textbf{158} & 24 & \textbf{86.8\%} & \textbf{79.8\%} & \textbf{83.2\%} \\
TruffleHog & 71 & 8 & 89.9\% & 35.9\% & 51.3\% \\
SecretFinder & 63 & 1{,}402 & 4.3\% & 31.8\% & 7.6\% \\
JSluice & 62 & 89 & 41.1\% & 31.3\% & 35.5\% \\
Titus & 40 & 11 & 78.4\% & 20.2\% & 32.1\% \\
JSMiner & 12 & 6 & 66.7\% & 6.1\% & 11.1\% \\
Nuclei & 7 & 0 & 100.0\% & 3.5\% & 6.8\% \\
Sensitive Discoverer & 7 & 8 & 46.7\% & 3.5\% & 6.6\% \\
Cariddi & 2 & 0 & 100.0\% & 1.0\% & 2.0\% \\
\midrule
\multicolumn{6}{@{}l}{\emph{Labelling reference (defined the ground truth; not a competing scanner)}}\\
\textit{Claude Opus 4.7 (oracle)} & 161 & 30 & \textit{84.3\%} & \textit{81.3\%} & \textit{82.8\%} \\
\bottomrule
\end{tabular}
\end{table}

Three results stand out. First, on this corpus of real production data
SecretSifter recovers 79.8\% of GT-198 (158 of 198) in default configuration,
which is 2.2 times the recall of the best static scanner (TruffleHog at 35.9\%), a
lead of 43.9 percentage points. Second, this recall does not come at the cost of
precision. SecretSifter emits 24 false positives at 86.8\% precision, giving an F1
of 83.2\%, the highest of any system in the set, marginally ahead even of the LLM
oracle (82.8\%, shown for reference only); its recall trails only that oracle. The contrast with the only other
broad-coverage extractor is stark: SecretFinder reaches a comparable 31.8\%
recall but emits 1{,}402 false positives, collapsing its precision to 4.3\%. The
difference is architectural:
SecretSifter's context-gated suppression cascade (Section~\ref{sec:func}) removes
framework directives, build-artifact hashes, and other noise before it becomes a
finding, through mechanisms such as a placeholder filter, a semantic key-name
requirement, over 90 forced-noise keys, and a 30-domain CDN blocklist; SecretFinder has no comparable stage, so generic UUID, 32-hex,
and base64 patterns matching webpack chunk hashes, build fingerprints, and
framework constants saturate its output. The narrow parser-based scanners (Titus
at 78.4\% precision, JSluice at 41.1\%) keep noise down but recover far less
(20.2\% and 31.3\% recall), so SecretSifter's F1 lead is structural: it holds
high recall together with high precision where the other tools sacrifice one for
the other. Third, the CryptoJS-AES encrypted-configuration class (37 of the 198 values)
shows why the decryptor matters: out of the box, pattern-only scanners see an
opaque \texttt{Salted\_\_} blob and flag none of the 37, whereas SecretSifter's
dedicated decryptor recovers all 37, the single largest component of its recall
lead. Precision and recall are measured on real production credentials that cannot be
redistributed, so the figures are not directly reproducible by third parties; a
de-identified reproducibility bundle (the per-value detection matrix and the
pairwise significance results) is provided with the extended evaluation
report~\cite{gorijala2026zenodo} so the scoring can be inspected
without exposing the confidential values. The full protocol, including the exact
per-tool configurations and the shared six-rule overlay (reproduced verbatim in
its appendix), the per-tool versions and sources, the significance
testing, and the detection-set overlap analysis, is documented in that
report~\cite{gorijala2026zenodo}.

Beyond these quantitative results, the tools differ qualitatively in where they
look and how they are deployed. Table~\ref{tab:qualitative} positions SecretSifter
against the eight evaluated scanners on license, observation surface, and
deployment. The distinguishing axis is the last column: only SecretSifter observes
secrets that exist solely after client-side JavaScript executes --- values in the
rendered DOM and in-memory state --- whereas the others inspect source bytes on
disk, in a repository, or on the wire.

\begin{table*}[t]
\centering
\footnotesize
\caption{Qualitative positioning of SecretSifter against the eight scanners
evaluated in Table~\ref{tab:results}. The eight comparison scanners are free and
open source; commercial use is permitted under each license, with copyleft
obligations for the GPL/AGPL tools (TruffleHog, SecretFinder, Cariddi) when the
tool is modified or, for AGPL, offered as a network service. SecretSifter's
detection core and its Burp and AI/MCP extensions are open source under the MIT
license, while its Chrome and desktop editions are distributed as packaged builds
under a proprietary end-user license (see Code metadata~C4). \emph{Rendered
runtime} marks whether a tool observes secrets that exist only after client-side
JavaScript executes (rendered DOM and in-memory state), as opposed to bytes on
disk or on the wire.}
\label{tab:qualitative}
\begin{tabular}{@{}p{2.0cm} l p{3.3cm} p{2.5cm} c@{}}
\toprule
Scanner & License & Observation surface & Deployment & \shortstack{Rendered\\runtime}\\
\midrule
\textbf{SecretSifter} & MIT / proprietary & Live HTTP traffic, JS-rendered runtime (CDP / page injection), and static assets & Burp extension, Chrome/Firefox, desktop app & \textbf{Yes}\\
\addlinespace
TruffleHog          & AGPL-3.0   & Git history and filesystem            & CLI (Go)          & No\\
SecretFinder        & GPL-3.0    & Static JavaScript from fetched URLs   & CLI (Python)      & No\\
JSluice             & MIT        & Static JavaScript (AST parse)         & CLI / Go library  & No\\
Titus               & Apache-2.0 & Files and repositories (static)       & CLI (Go)          & No\\
JS Miner            & Apache-2.0 & Static JS/JSON in proxied traffic     & Burp extension    & No\\
Nuclei              & MIT        & HTTP responses via templates          & CLI (Go)          & No\\
Sensitive Discoverer& Apache-2.0 & HTTP messages in proxied traffic      & Burp extension    & No\\
Cariddi             & GPL-3.0    & Crawled endpoints and JavaScript      & CLI (Go)          & No\\
\bottomrule
\end{tabular}
\end{table*}

\subsection{Engineering contribution and reuse}
The software contribution, independent of the benchmark, is a reuse architecture.
A single detection core (\texttt{SecretScanner} and one rule library) drives
three deployment models, an in-path proxy, an in-browser extension, and a
standalone crawler, through thin capture adapters rather than three forked
codebases; the detection logic is written once and reused across proxy,
in-browser, and standalone contexts. The open-source artifact is this Java core
and its Burp capture layer (C2); the Chrome and desktop ports reuse the same
design as packaged distributions, so the directly inspectable and extensible
contribution is the open core rather than all three editions' capture code. The detection core ships with a suite of over 90 unit
tests (\texttt{PatternsTest} and \texttt{SecretScannerTest}) covering the rule
library and the entropy and decryption logic, so the
behavior other deployments reuse is regression-tested. The same core is exposed for automation
through a Bearer-authenticated REST and MCP interface, so an external LLM agent or
CI pipeline can run scans, manage rules, and drive cooperative triage with no GUI.
This makes SecretSifter reusable infrastructure rather than a single-use product:
its detection core and export interface served as the measurement harness for the
companion study, and the three packaged editions plus the open InsecureShield
reference target let other researchers reproduce and extend runtime
secret-exposure experiments without a commercial proxy licence. The tool also
supports a workflow open tooling did not previously offer, continuous runtime
secret monitoring during penetration-testing and AppSec engagements. It is distributed through the Chrome Web Store and public GitHub repositories.

\subsection{Limitations}
Several limitations bound these results. The benchmark is drawn from a single
organization's applications and is not a cross-industry population estimate. The
tool under evaluation was developed by the author; the LLM-based benchmark
construction and cross-vendor validation are the mitigations, and the competing
interest is declared. A related coupling must be stated plainly: the detection
core and the benchmark derive from the same organizational context, so several
detection categories, notably the Azure APIM and CryptoJS-AES classes that
dominate GT-198, were developed with prior awareness of the credential types
present in this environment. The recall figures should therefore be read as
in-context detection performance rather than an out-of-distribution
generalization estimate, and the rule-extensible baselines, which received only
the six-rule overlay, do not carry the same environment-specific tuning. The
ground truth is anchored to LLM extraction (Claude Opus 4.7) with cross-vendor
validation (GPT-5.5) and a manual analyst pass, rather than to an exhaustive
independent human relabelling of every bundle, so a client-exposed value that
both the model and the analyst overlooked would not appear in GT-198 and would
not penalise any scanner; the recall denominator is a high-quality but not
provably complete ground truth. Because GT-198 contains real credentials it cannot be
redistributed, so the precision and recall figures are not directly reproducible
by third parties; the InsecureShield demo target (Section~\ref{sec:examples}) is
provided as the reproducible artifact instead. Because the corpus is one
organization's production applications, the false-positive counts reflect that
traffic's noise profile and may differ on other stacks. Finally, the reported recall reflects the deterministic
detection stage; the optional triage layer is excluded from these figures.

%% ======================= 5 =======================
\section{Conclusions}
SecretSifter implements runtime secret detection as a deployable, extensible
engine that closes a structural gap left by repository-time scanners. It is
delivered as three editions (Burp, Chrome, and a standalone desktop application)
over one shared detection core, and on a benchmark of real production
values (GT-198), labelled independently of the tools under test, it achieves the
highest F1 of any system evaluated in default configuration, combining high recall
with 86.8\% precision. Future work includes
multi-organization benchmarking, expanded decryptor coverage, and deeper CI
integration through the REST API. The shared detection core is open-source (MIT) and publicly available at the two
repositories listed in C2; the Chrome and desktop editions are packaged builds of
that same core.

\section*{Declaration of competing interest}
The author is the developer of SecretSifter, one of the scanners referenced in
the evaluation. The benchmark and comparison protocol were constructed
independently of tool output (see the companion study), and this relationship is
disclosed.

\section*{Acknowledgements}
The author thanks the organizers and attendees of BSides San Antonio and BSides
Buffalo for feedback on early demonstrations of SecretSifter.

%% ======================= AI declaration =======================
\section*{Declaration of generative AI and AI-assisted technologies in the writing process}
During the preparation of this work the author used generative AI tools to draft and
edit the manuscript text and format the LaTeX source. After using these tools, the
author reviewed and edited the content as needed and takes full responsibility for
the content of the publication. No generative AI was used to produce the research,
the software, the benchmark, or the analysis reported here.

%% ======================= References =======================

\end{document}